\documentclass[aps,prl,twocolumn]{revtex4}
\usepackage{graphicx}
\usepackage{dcolumn}
\usepackage{amsmath}
\usepackage{amssymb}
\usepackage{subfigure,amsmath,verbatim,moreverb,bm}
\def\be{\begin{equation}}
\def\ee{\end{equation}}
\def\ber{\begin{eqnarray}}
\def\eer{\end{eqnarray}}

\def\rv{{\bf r}}
\def\xv{{\bf x}}

\begin{document}
\title{A variational approach to London dispersion interactions without density distortion}
\author{Derk P. Kooi}
\affiliation{Department of Theoretical Chemistry and Amsterdam Center for 
Multiscale Modeling, Faculty of Science, Vrije Universiteit Amsterdam, The Netherlands}
\author{Paola Gori-Giorgi}
\affiliation{Department of Theoretical Chemistry and Amsterdam Center for 
Multiscale Modeling, Faculty of Science, Vrije Universiteit Amsterdam, The Netherlands}

\date{\today}

\begin{abstract}
We introduce a class of variational wavefunctions that capture the long-range interaction between neutral systems (atoms and molecules) without changing the diagonal of the density matrix of each monomer. The corresponding energy optimization yields explicit expressions for the dispersion coefficients in terms of the ground-state pair densities of the isolated systems, 
providing a clean theoretical framework to build new approximations in several contexts. As the individual monomer densities are kept fixed, we can also unambiguously assess the effect of the density distortion on London dispersion interactions: for example, we obtain virtually exact dispersion coefficients between two hydrogen atoms up to $C_{10}$, and relative errors below $0.2\%$ in other simple cases. 
\end{abstract}
\maketitle


Within the Born-Oppenheimer approximation, non-relativistic quantum mechanics of Coulomb systems (atoms, molecules) predicts that everything is attracted to everything, since it is always possible to find a relative orientation that lowers the total energy of two neutral molecules separated by a large distance $R$ by at least a term $-C_6R^{-6}$, with $C_6>0$ \cite{LieThi-PRA-86,Kap-Book-06}. This universal attractive interaction between quantum Coulomb systems, often called London dispersion interaction, is very small compared to the energy scale of a typical chemical (covalent) bond. Yet, it is omnipresent in chemistry, biology, and physics, determining protein folding, the structure of DNA, the physics of layered materials, just to name a few examples. An accurate, efficient, and fully non-empirical treatment of these forces remains an open challenge in many respects, and it is the focus of several on-going research efforts \cite{Gri-WIRCMS-11,BurMaySum-JCP-11,DobGou-JPCM-12,GriHanBra-CR-16,VydVoo-JCP-10,FerDiSAmb-PRL-15,Bec-JCP-13}.

The physical origin of London dispersion interactions is often discussed in terms of ``coupling between instantaneous dipoles'' \cite{Kap-Book-06,Sto-book-96}, which  are actually correlations in the interfragment part of the pair density (the diagonal of the two-body reduced density matrix), as illustrated, e.g., in \cite{GriBae-JCP-06}. The textbook treatment is based on perturbation theory \cite{Kap-Book-06,Sto-book-96,MarKesTer-Book-71}, in which excitations on both monomers are coupled, which allows to rewrite the dispersion coefficients in terms of dynamical polarizabilities of the individual systems. 
In this Letter, inspired by the work of Lieb and Thirring (LT) \cite{LieThi-PRA-86}, we attack the dispersion problem from a variational perspective, using only ground-state properties of the monomers. We construct a class of variational wavefunctions for the long-range interaction between neutral Coulomb systems that are explicitly forbidden to deform the diagonal of the full spatial density matrix (and thus the density) of each individual monomer. This constraint has several advantages: first of all, it is conceptually appealing, as dispersion becomes a monomer-monomer interaction accompanied by a change in kinetic energy only, since all the remaining intra monomer interactions are kept unchanged. Secondly, it highly simplifies the optimization procedure, which can be implemented in a very efficient way and leads to expressions for the dispersion coefficients in terms of the ground-state pair densities of the individual monomers, giving a neat framework to build new approximations.  For example, in a density functional theory (DFT) setting, one can use approximations for the exchange-correlation holes of the fragments, providing justification and a route for improving models such as the exchange-hole dipole moment (XDM) \cite{BecJoh-JCP-07}. It also provides a prescription to obtain atomic dispersion coefficients inside molecules and solids, which are a crucial ingredient in the construction of atomistic force fields and dispersion corrections to DFT calculations (DFT+D).

One could easily object that this construction cannot give the exact answer. In a widely quoted paper, Feynman \cite{Fey-PR-39} wrote that van der Waals' forces arise from charge distributions with higher concentration between the nuclei, with a permanent dipole moment of order $R^{-7}$ being induced on each atom, such that {\em  ``each nucleus is attracted by the distorted charge distribution of its own electrons''} \cite{Fey-PR-39}. Feynman's conclusion was proven by using the electrostatic Helmann-Feynman theorem for both atoms and molecules \cite{Hun-JCP-90}, and has been confirmed with various accurate calculations \cite{AllToz-JCP-02,StrKumCor-JCP-11}. Therefore we know that a small density distortion must be there. Nonetheless, it has been observed several times that very accurate dispersion coefficients can be obtained with methods that describe very poorly this density distortion, including cases in which the electrostratic Helmann-Feynman theorem gives zero force at order $R^{-7}$ \cite{HirEli-JCP-67}. The issue has been often debated in the literature, and there is the general feeling that the density distortion, which actually requires expensive calculations to be accurately captured, should have very little influence on dispersion energetics, which is our goal here. Moreover, our construction provides (if combined with accurate pair densities for the monomers), a rigorous upper bound to the best possible variational interaction energy when the density distortion is explicitly forbidden, allowing us to assess its importance in an unambiguous manner, which is {\em per se} also conceptually interesting. 

{\it The wavefunction for $N=2$ --} For illustrative purposes, we discuss first the case of two systems $A$ and $B$, each with $N_A=N_B=1$ electron, generalizing it right after to the many-electron case. The two atoms are separated by a large distance $R$, and our wavefunction reads 
\begin{equation}
	\Psi_R(\rv_1,\rv_2)=\Psi_0^A(\rv_1)\Psi_0^B(\rv_2)\sqrt{1+J_R(\rv_1,\rv_2)},
	\label{eq:PsiRN2}
\end{equation}
where $\rv_1$ and $\rv_2$ are electronic coordinates centered in nucleus $A$ and nucleus $B$, respectively, $\Psi_0^{A(B)}(\rv)$ is the ground-state wavefunction of system $A$ (or $B$) alone, and $J_R$, with $|J_R(\rv_1,\rv_2)|\ll 1$, preserves the unperturbed densities at all orders in $R^{-1}$ via the constraints
\begin{align}
	& \int d\rv_1 |\Psi_0^{A}(\rv_1)|^2 J_R(\rv_1,\rv_2)=0 \qquad \forall\; \rv_2 \label{eq:constrJ1} \\
	& \int d\rv_2 |\Psi_0^{B}(\rv_2)|^2 J_R(\rv_1,\rv_2)=0 \qquad \forall \;\rv_1,
\label{eq:constrJ2}
\end{align}
which are imposed with an expansion in terms of one-electron functions $b_i^{A}(\rv)$ and $b_i^{B}(\rv)$ and variational parameters $c_{ij,R}$,
\begin{equation}
	\label{eq:JN2}
	J_R(\rv_1,\rv_2)=\sum_{ij}c_{ij,R} b_i^A(\rv_1)b_j^B(\rv_2),
\end{equation}
where
\begin{equation}
b_i^A(\rv)=f_i^A(\rv)-\int \frac{\rho_0^A(\rv)}{N_A}f_i^A(\rv)d\rv
\end{equation}
(and similarly for $b_j^B$), with $\rho_0^{A(B)}(\rv)=|\Psi_0^{A(B)}(\rv)|^2$. The $f_i^{A(B)}(\rv)$'s can be chosen in different ways, with the choice influencing the convergence. Our wavefunction does not include antisymmetrization between the electron on $A$ and the one on $B$, which affects the asymptotic (large $R$) interaction energy only with terms that vanish exponentially with $R$. 
This wavefunction correlates the positions of the two electrons in the two different atoms without changing their one-electron densities. 

We now consider the full hamiltonian of the problem, $\hat{H}=\hat{H}_A(\rv_1)+\hat{H}_B(\rv_2)+\hat{H}_{\rm int}(\rv_1,\rv_2)$, where $\hat{H}_{A}$ and $\hat{H}_{B}$ are the hamiltonians of each isolated atom and $\hat{H}_{\rm int}$ contains all the interactions between the particles in $A$ (electron and nucleus) with those in $B$, and evaluate its expectation value on our wavefunction, focussing on the interaction energy $E_{\rm int}=E_0^{AB}-E_0^A-E_0^B$. 
Thanks to the density constraint of Eqs.~\eqref{eq:constrJ1}-\eqref{eq:constrJ2}, there is no change in the expectation value of the nuclear-electron potential, implying that the interaction energy is entirely determined by the change in kinetic energy $\hat{T}$ with respect to the one of the isolated atoms,
 \begin{equation}
 \Delta\,T(\{c_{ij,R}\})=\langle\Psi_R|\hat{T}|\Psi_R\rangle-T_0^A-T_0^B,
 \end{equation} plus the expectation value of $\hat{H}_{\rm int}$, which splits automatically into the sum of an electrostatic part $W_{\rm int,0}(R)$ and the correlation part $W_{\rm int,c}(\{c_{ij,R}\},R)$,
 \begin{align}
& W_{\rm int,0}(R)=\langle\Psi_0^{A}\Psi_0^{B}|\hat{H}_{\rm int}|\Psi_0^{A}\Psi_0^{B}\rangle\,,  \\
& W_{\rm int,c}(\{c_{ij,R}\},R)=\int \hat{H}_{\rm int}\,|\Psi_0^{A}|^2\,|\Psi_0^{B}|^2\,J_R\, d\rv_1d\rv_2\,, \\
&  E_{\rm int}(R)=\Delta\,T(\{c_{ij,R}\})+W_{\rm int,0}(R)+W_{\rm int,c}(\{c_{ij,R}\},R).
 \label{eq:Eint}
\end{align}
As usual \cite{Kap-Book-06}, we perform a multipolar expansion of $\hat{H}_{\rm int}$,
\begin{equation}
	\label{eq:multip}
	\hat{H}_{\rm int}= \frac{\hat{H}_{dd}}{R^3} + \frac{\hat{H}_{dq}+\hat{H}_{qd}}{R^4} + \frac{\hat{H}_{qq}}{R^5}+...,
\end{equation}
where $d$ stands for dipole, $q$ for quadrupole, etc.
The variational problem then takes a simplified form, as detailed in the supplementary material \cite{sup}, with each order $R^{-2n}$ in the energy having its optimal $c_{ij,R}=c_{ij}\,R^{-n}$. The physics described by this wavefunction can be understood by looking at the leading order: our wavefunction gives a $\Delta\,T(\{c_{ij}\})$ which is positive (as it costs kinetic energy to correlate the electrons) and {\em quadratic} in the variational parameters $c_{ij}$, while $W_{\rm int,c}(\{c_{ij}\},R)$ is linear in the $c_{ij}$, negative (as correlation occurs exactly to lower the interaction term), and has a prefactor $R^{-3}$ from \eqref{eq:multip}. This guarantees that there are always optimal variational parameters $c_{ij}$, all proportional to $R^{-3}$, which minimize $E_{\rm int}$, whose minimum becomes the familiar $-C_6/R^6$. This structure repeats  at each order in $1/R$.
In LT \cite{LieThi-PRA-86} and subsequent works in a similar spirit,\cite{AnaSig-CPAM-17,AnaLew-arXiv-18} the energy cost to correlate the electrons, quadratic in the variational parameters, comes from the full hamiltonian of the monomers, while with our construction we force it to be of kinetic energy origin only. 

Solving the variational equations leads to explicit expressions for all the even dispersion coefficients. For example, $C_6$ is given by
\begin{equation}
	\label{eq:C6}
	C_6=\frac{1}{2}\mathbf{w}^T L^{-1} \mathbf{w},
\end{equation} 
where the vector $\mathbf{w}$ has contracted indices $I=ij$,
\begin{equation}
	w_{ij}=\sum_{e}h_e \int e|\Psi_0^A(\rv)|^2b_i^A(\rv)d\rv\int e|\Psi_0^B(\rv)|^2b_j^B(\rv)d\rv,
\end{equation}  
with $e=x,y,z$, and $h_e=(1,1,-2)$ when the bond is along the $z$-axis. The matrix $L$ has two contracted indices $I=ij$ and $K=kl$,
\begin{equation}
	L_{ij,kl}=\frac{1}{4}\left(\tau_{ik}^AS_{jl}^B+S_{ik}^A\tau_{jl}^B\right),
\end{equation}
with
\begin{align}
	& \tau_{ij}^A=\int |\Psi_0^A(\rv)|^2 \nabla b_i^A(\rv)\cdot \nabla b_j^A(\rv)\, d\rv, \label{eq:tauij}\\
	& S_{ij}^A=\int |\Psi_0^A(\rv)|^2 b_i^A(\rv) b_j^A(\rv)\, d\rv, \label{eq:Sij}
\end{align}
and similarly for $B$. Analogous expressions hold for all the even dispersion coefficients (see \cite{sup}).

\begin{figure}
\includegraphics[width=7.5cm]{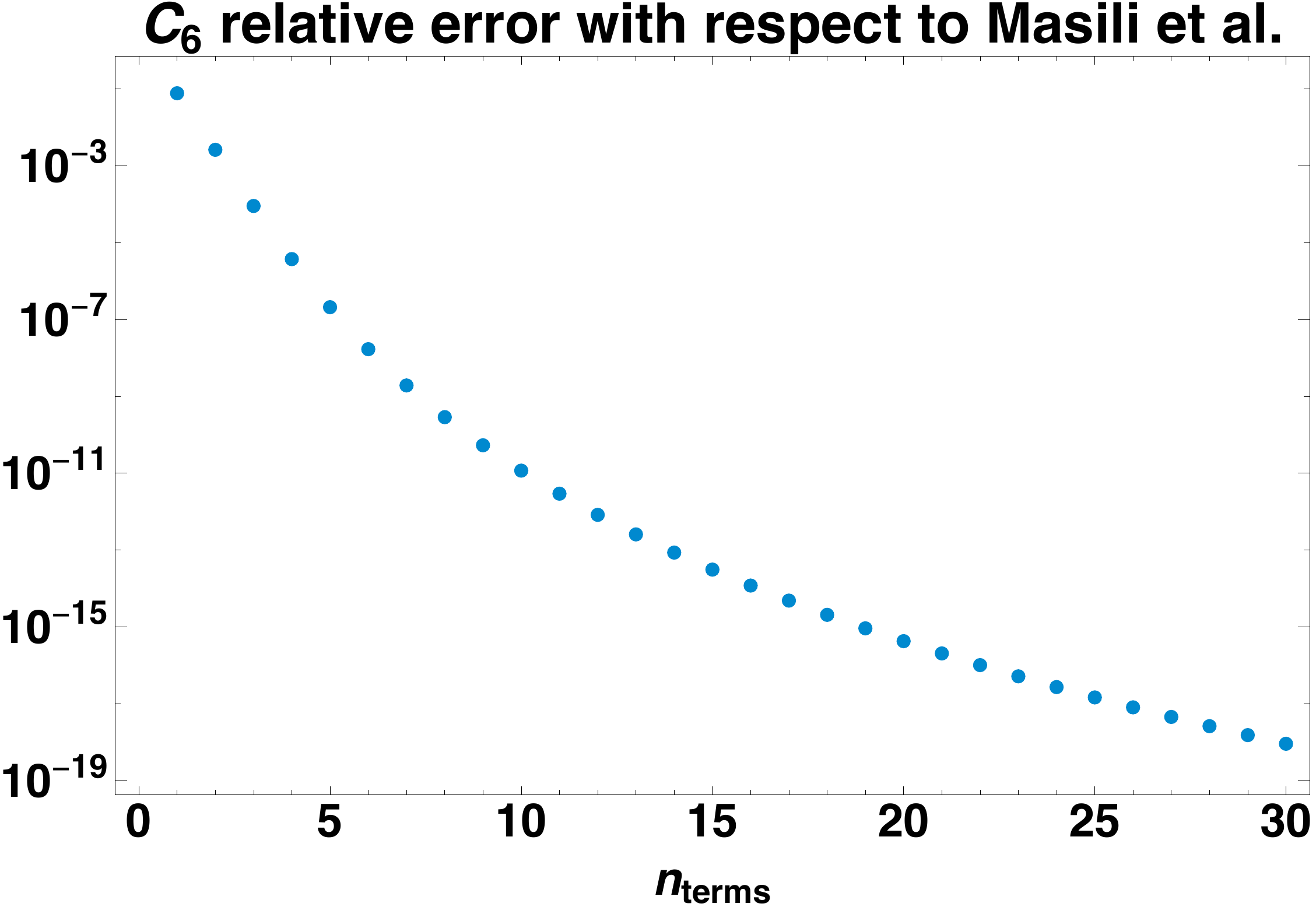}
\caption{Relative error on $C_6$ between two H atoms with increasing number of terms in Eq.~\eqref{eq:JN2} relative to the reference result \cite{MasGen-PRA-08}.}
\label{fig:convH}
\end{figure}

We have started with a very simple choice for the $f_i^{A(B)}(\rv)$'s (just powers of the distance $r$ from the nucleus times spherical harmonics \cite{sup}), finding that the convergence is already fast, as shown in Fig.~\ref{fig:convH} for the case of $C_6$ for two H atoms. The values of $C_6$, $C_8$ and $C_{10}$ for the same case are reported in Table~\ref{tab:CnH}, where they are compared with reference data from Ref.~\citenum{Tha-JCP-88}. In the supplementary material \cite{sup}, we also report even second-order coefficients up to $C_{30}$, which also agree within numerical accuracy with those of Ref.~\citenum{Tha-JCP-88}. From Table~\ref{tab:CnH} we see that our wavefunction yields essentially the exact lowering in energy due to dispersion without any distortion of the density of each individual atom. 


\onecolumngrid

\begin{table}
\begin{tabular}{l|c|c|c}
\hline \hline
       &  Thakkar \cite{Tha-JCP-88} & Masili et al.\cite{MasGen-PRA-08}& This work (30 terms)\\
\hline
	  $C_6$ & $6.4990267054058405\times 10^0$ & $6.4990267054058393\times 10^0$ & $6.4990267054058\mathbf{393}\times 10^0$ \\
	  $C_8$ & $1.2439908358362235\times 10^2$ & n.a. & $1.243990835836223\mathbf{4}\times 10^2$ \\
	  $C_{10}$ & $3.2858284149674217\times 10^3$ & n.a. & $3.2858284149674217\times 10^3$ \\
\hline \hline
\end{tabular}
\caption{Dispersion coefficients $C_6$, $C_8$ and $C_{10}$ between two H atoms computed with 30 terms in the variatonal expression of Eqs.~\eqref{eq:PsiRN2}-\eqref{eq:Sij}, compared to the reference data of Thakkar \cite{Tha-JCP-88} and Masili, et al.\cite{MasGen-PRA-08}. See also supplementary material \cite{sup}. }  
\label{tab:CnH}
\end{table}
\twocolumngrid

{\it The many-electron case --} We now turn to the many-electron case, by considering two systems with $N_A$ and $N_B$ electrons, for which our wavefunction reads
\begin{equation}
	\Psi_R(\underline{\xv}_A,\underline{\xv}_B)=\Psi_0^A(\underline{\xv}_A)\Psi_0^B(\underline{\xv}_B)\sqrt{1+\sum_{i\in A,j\in B}J_R(\rv_i,\rv_j)},
\end{equation}
where $\xv_{A(B)}$ indicates all the electronic coordinates (including spins) of the electrons in system $A$ (or $B$), and, again, $\Psi_0^{A(B)}$ is the ground state wavefunction of system $A$ (or $B$) alone. We choose $J_R$ such that the diagonal of the full many-body density matrix of each system is not allowed to deviate from its ground-state expression:
\begin{align}\label{eq:PsiA2}
&	\int |\Psi_R(\underline{\xv}_A,\underline{\xv}_B)|^2 d\xv_A= |\Psi_0^B(\underline{\xv}_B)|^2\qquad \forall\, \xv_B \\
&   \int |\Psi_R(\underline{\xv}_A,\underline{\xv}_B)|^2 d\xv_B= |\Psi_0^A(\underline{\xv}_A)|^2\qquad \forall\, \xv_A,
\label{eq:PsiB2}
\end{align}
which is enforced by choosing for $J_R$ the same form as Eq.~\eqref{eq:JN2}, i.e., by imposing that
\begin{align}
	& \int \rho_0^A(\mathbf{r}_{i_A})J_R(\mathbf{r}_{i_A},\mathbf{r}_{j_B}) \mathrm{d} \mathbf{r}_{i_A}= 0\qquad \forall\; \mathbf{r}_{j_B} \\
	& \int \rho_0^B(\mathbf{r}_{j_B})J_R(\mathbf{r}_{i_A},\mathbf{r}_{j_B}) \mathrm{d} \mathbf{r}_{j_B}= 0\qquad \forall\; \mathbf{r}_{i_A},
\end{align}
with $\rho_0^{A(B)}$ the ground-state one-electron densities of the two systems. Again, it is only the interfragment pair density and the off-diagonal elements of the first-order reduced density matrix that are allowed to change to lower the energy, implying that also in the many-electron case the interaction energy is given by Eq.~\eqref{eq:Eint}, since the expectation value of the electron-electron interaction inside each monomer is kept unchanged by virtue of Eqs.~\eqref{eq:PsiA2}-{\eqref{eq:PsiB2}. The variational minimization of the interaction energy is similar to the $N=2$ case. For example, $C_6$ takes the same form as Eq.~\eqref{eq:C6}, where now $\mathbf{w}$ and $L$ read
\begin{align}
& w_{ij} = \sum_{e=x,y,z} h_e (d_i^A+D_i^A)(d_j^B +D_j^B), \\
& L_{ij,kl} = \frac{1}{4}\left(\tau_{ik}^A(S^B_{jl}+P^B_{jl})+(S^A_{ik}+P^A_{ik})\tau_{jl}^B\right), 
\end{align}
where (with the same expressions for $B$)
\begin{align}
 d_i^A & = \int \mathrm{d} \mathbf{r}_{1_A} \rho_0^A(\mathbf{r}_{1_A})b_i^A(\mathbf{r}_{1_A}) e_{1_A}, \\
 D_i^A & = \int \mathrm{d} \mathbf{r}_{1_A} \int \mathrm{d} \mathbf{r}_{2_A} P_0^A(\mathbf{r}_{1_A}, \mathbf{r}_{2_A}) b_i^A(\mathbf{r}_{2_A}) e_{1_A}, \\
  P_{ij}^A & =  \int \mathrm{d} \mathbf{r}_{1_A} \int\mathrm{d} \mathbf{r}_{2_A} P_0^A(\mathbf{r}_{1_A}, \mathbf{r}_{2_A}) b_i^A(\mathbf{r}_{1_A}) b_j^A(\mathbf{r}_{2_A}),
\end{align}
and where $\tau_{ij}^{A(B)}$ and $S_{ij}^{A(B)}$ are defined in Eqs.~\eqref{eq:tauij}-\eqref{eq:Sij}. 
We see that the dispersion coefficients $C_n$ for the many-electron case become explicit functionals of the spin-summed ground-state pair densities $P_0^A(\mathbf{r}_{1_A}, \mathbf{r}_{2_A})$ and $P_0^B(\mathbf{r}_{1_B}, \mathbf{r}_{2_B})$ of the two monomers,
$C_n=C_n[P_0^A,P_0^B]$.
The variational solution for the optimal $c_{ij}$ can also be turned into a Sylvester equation, which can be solved in an efficient way \cite{BarSte-ACM-72,sup}. 

Being derived from a wavefunction, the expression for the interaction energy is variational (i.e., it provides a lower bound to the exact $C_6$) as long as we use exact (or very accurate) pair densities for the monomers. As an example, with the same simple choice for the $f_i^{A(B)}(\rv)$ used for two H atoms, we have computed $C_6$ for He-He using the accurate variational wavefunction for the He atom of Freund, Huxtable and Morgan \cite{FreHuxMor-PRA-84} (see also Ref.~\citenum{MirUmrMorGor-JCP-14}). The convergence is again fast, very similar to the one of Fig.~\ref{fig:convH}, yielding $C_6=1.458440$ with an error of 0.17\% with respect to the accurate value 1.460978 of Ref.~\citenum{YanBabDal-PRA-96}. For the He-H case we have a similar error of 0.15\% with respect to Ref.~\citenum{YanBabDal-PRA-96}. Whether these residual errors are due to the reduced variational freedom of our wavefunction or to inaccuracies in the multiple moment integrals of the correlated He pair density of Ref.~\citenum{FreHuxMor-PRA-84} is for now an open question, but we see that the results can be rather accurate, at a computational cost which is determined by the individual monomer calculations. 
This opens many possibilites, as our expressions can be coupled with any wavefunction method for the monomers, provided it gives access to the second-order reduced density matrices and their integrals with the $f_i^{A(B)}(\rv)$, whose optimal choice needs to be assessed in a systematic way.

\begin{figure}
\includegraphics[width=7.5cm]{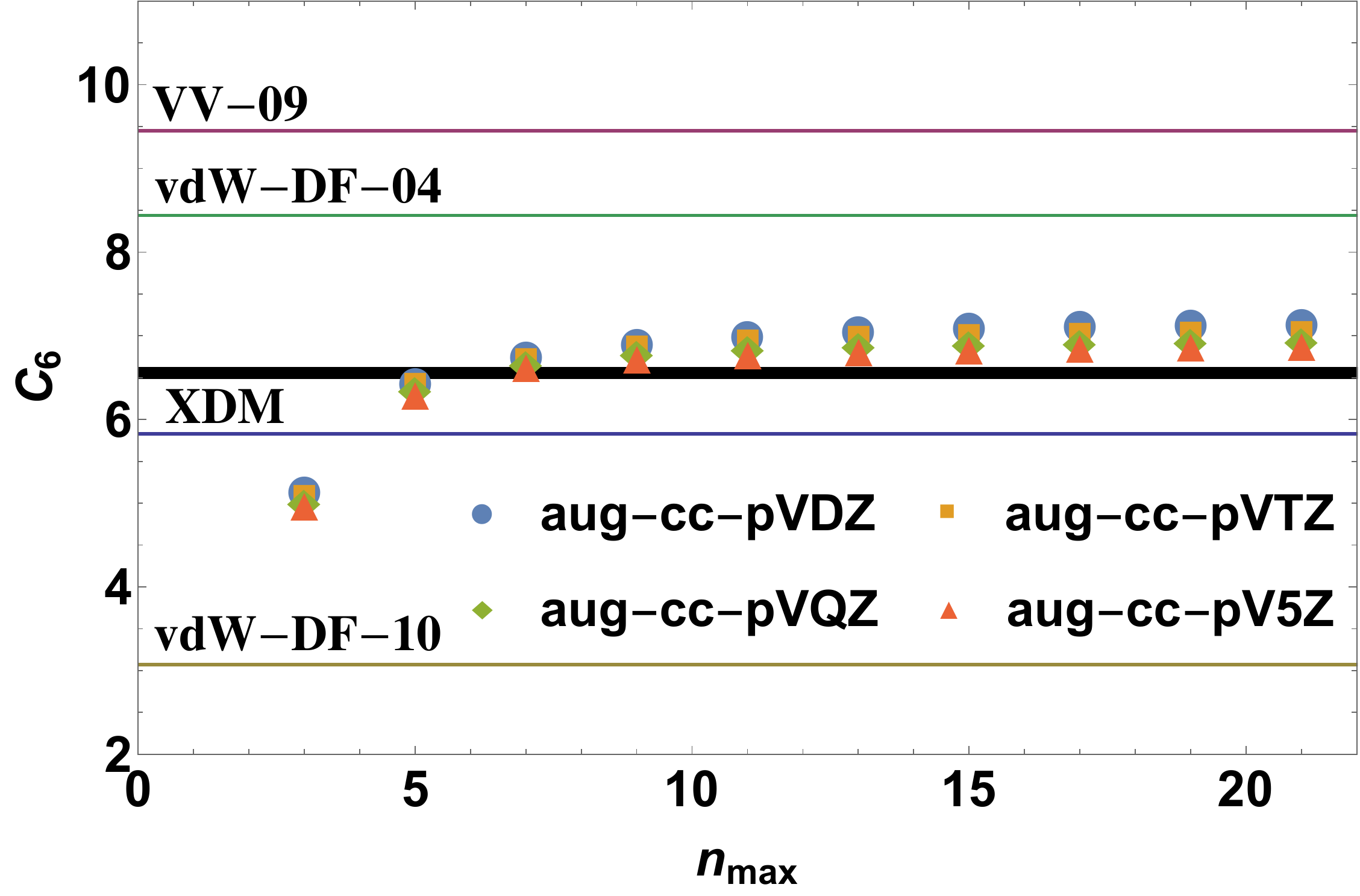}
\caption{Convergence of the $C_6$ dispersion coefficient with the number of functions $f_i^{A(B)}(\rv)$ for Ne-Ne using Hartree-Fock pair densities of the individual atoms in different basis sets. HF calculations were carried out using Gaussian 16\cite{Gaussian16} and the arbitrary order multipole moments were calculated using HORTON 3.\cite{Horton3} Accurate value (thick line) from Ref.~\citenum{ThaHetWor-JCP-92}. Values for vdW-DF-04/10 and VV-09 are from Ref.~\citenum{VydVoo-PRA-10}, while the value for XDM was taken from Ref.~\citenum{BecJoh-JCP-07}. vdW-DF-09 was not included, since it is parametrized based on this dispersion coefficient.}
\label{fig:Ne}
\end{figure}

{\it  Lower-level approximations --} We can of course use pair densities from lower level theories, which opens another whole realm of possibilities, bearing in mind that the expressions will be no longer variational for the interaction energy. For example, in Fig.~\ref{fig:Ne} we show the result for $C_6$ for Ne-Ne using the Hartree-Fock (HF) pair density, including the basis set dependence. We see that the error is around 5\%, as HF is a good approximation for the Ne atom, and that we are below the exact interaction energy. It is clear that a detailed study of the basis set dependence is also needed, which should be coupled to an optimal choice for the  $f_i^{A(B)}(\rv)$. The use of Kohn-Sham orbitals from different approximate functionals to determine the exchange pair densities is also worth to be investigated. 

Within DFT, one could make even simpler approximations for the exchange-correlation holes $h_{xc}^{A(B)}$ of the monomers, transforming the dispersion coefficients into density functionals,
\begin{equation}
	C_n[P_0^A,P_0^B]=C_n[\rho_0^A,h_{xc}^{A},\rho_0^B,h_{xc}^{B}]\to C_n[\rho_0^A,\rho_0^B].
\end{equation}
For example, the Becke-Roussel \cite{BecRou-PRA-89} exchange-hole model could be used to estimate higher moments of the exchange pair density \cite{GorAngSav-CJC-09}, in a spirit similar to the XDM model \cite{BecJoh-JCP-07}. The main difference is that now the polarizability of the monomers is no longer needed, as the expression for the dispersion coefficients only requires the exchange-correlation holes.

An important ingredient in atomistic force fields models and DFT+D are the dispersion coefficients $C_{n,ab}$ between pairs of atoms $a\,b$, with $a$ in system $A$ and $b$ in system $B$, which are usually computed from atomic polarizabilities. Our variational wavefunction can provide a framework to define and compute them. One should obtain atomic volumes $\Omega_a^A$ and $\Omega_b^B$ using one of the various definitions (e.g. Hirshfeld partitioning \cite{Hir-TCA-77}), and then obtain the pair dispersion coefficients using, order by order, the energy density coming from our wavefunction with the $c_{ij}$ optimized for the $AB$ interaction, restricting the integration on the atomic volumes. For example for $C_6$ (with $J$ below being the optimized factor at leading order),
\begin{equation}
	\begin{split}
&	C_{6,ab} = \frac{1}{2} \int_{\Omega_a^A} \mathrm{d} \mathbf{r}_{1_A} \int_{\Omega_b^B}  \mathrm{d} \mathbf{r}_{1_B} \hat{H}_{dd}(\mathbf{r}_{1_A}, \mathbf{r}_{1_B}) \Big( \rho_0^A(\mathbf{r}_{1_A}) \rho_0^B(\mathbf{r}_{1_B})\\
	& \times J(\mathbf{r}_{1_A},\mathbf{r}_{1_B}) +  \int \mathrm{d} \mathbf{r}_{2_A} P_0^A(\mathbf{r}_{1_A},\mathbf{r}_{2_A}) \rho_0^B(\mathbf{r}_{1_B}) J(\mathbf{r}_{2_A},\mathbf{r}_{1_B})\\
&+  \int \mathrm{d} \mathbf{r}_{2_B}  \rho_0^A(\mathbf{r}_{1_A}) P_0^B(\mathbf{r}_{1_B},\mathbf{r}_{2_B}) J(\mathbf{r}_{1_A},\mathbf{r}_{2_B}) \\
&+ \int \mathrm{d} \mathbf{r}_{2_B} \mathrm{d} \mathbf{r}_{2_A}  P_0^A(\mathbf{r}_{1_A},\mathbf{r}_{2_A})  P_0^B(\mathbf{r}_{1_B},\mathbf{r}_{2_B}) J(\mathbf{r}_{2_A},\mathbf{r}_{2_B}) \Big)
	\end{split}
\end{equation}
and similarly for the higher order coefficients. 

{\it Conclusions --} In conclusion, the class of variational wavefunctions we have introduced here provides a neat theoretical framework to build new approximations to tackle dispersion interactions, which are reduced to a competition between kinetic energy and monomer-monomer interaction only. Although this class of variational wavefunctions violates the virial theorem (similarly to second-order perturbation theory), and although we know that for the exact wavefunction the density must be distorted, the results can be very accurate (or even exact), with a computational efficiency that is promising for the many-electron case. Moreover, we know what is being approximated, so that one could also devise a scheme in which the monomer (pair) densities are relaxed in a second step, or perturbatively, if this is needed. The realm of possibilities open is vast, ranging from the use of various wavefunction methods to compute the monomer pair densities or approximations thereof, to the definition and calculation of atomic-pair dispersion coefficients. The influence of the level of theory on the calculation of the monomer pair densities, the basis set, and different choices for the $f_i^{A(B)}(\rv)$ need all to be investigated. 
We have also shown that exact dispersion coefficients up to $C_{10}$ between two hydrogen atoms can be obtained without introducing any density distortion, which is interesting by itself. The framework can be also generalised to many monomers.

\noindent{\it Acknowledgements --} 
Financial support from European Research Council under H2020/ERC Consolidator Grant corr-DFT (Grant Number 648932) and the Netherlands Organisation for Scientific Research under Vici grant 724.017.001 is acknowledged.

\bibliography{../references/dispersion_bib.bib}

\end{document}


\title{A variational approach to London dispersion interactions without density distortion}
\author{Derk P. Kooi}
\affiliation{Department of Theoretical Chemistry and Amsterdam Center for 
Multiscale Modeling, Faculty of Science, Vrije Universiteit Amsterdam, The Netherlands}
\author{Paola Gori-Giorgi}
\affiliation{Department of Theoretical Chemistry and Amsterdam Center for 
Multiscale Modeling, Faculty of Science, Vrije Universiteit Amsterdam, The Netherlands}

\date{\today}

\maketitle

\section{Derivation of the equations for the  many-electron case}
The complete derivation of the expressions for the dispersion coefficients can be divided into three parts:
\begin{itemize} 
\item we first show explicitly that our variational wavefunctions preserve the many-body densities (diagonal of the full density matrix) of each monomer; 
\item
we then obtain the expectation values of the kinetic energy and of the monomer-monomer interaction hamiltonian; 
\item we finally perform a minimisation of the total energy, which splits order by order once we make a multipolar expansion of the interaction hamiltonian.
\end{itemize}
We denote the two neutral systems (fragments, monomers) by $A$ and $B$, having $N_A$ and $N_B$ electrons respectively, and nuclei fixed at their ground-state geometry.  Their centers are
separated by a large distance $R$ (the precise definition of the centers does not matter formally; they can be, e.g., the barycenters of nuclear charges). By $\underline{\xv}_{A/B} = \{ \xv_{1_{A/B}}, \dots \xv_{N_{A/B}}\}$ we denote the set of the (spatial and spin) coordinates of the electrons in fragment $A/B$, measured with respect to their respective centers. Our variational ansatz for the wavefunction consists of the product of the two ground-state wavefunctions of fragment $A$ and $B$ ($\Psi^A_0(\underline{\xv}_A) \Psi^B_0(\underline{\xv}_B)$) times a correlation (Jastrow) factor in the form of a square root, which contains the two-body correlation function $J_R(\rv_{i},\rv_{j})$,
\begin{equation}
\Psi_R(\underline{\xv}_A, \underline{\xv}_B) =  \Psi_0(\underline{\xv}_A, \underline{\xv}_B)\sqrt{1+ \sum_{i \in A, j \in B} \correltwo_R(\rv_{i}, \rv_j)}= \Psi^A_0(\underline{\xv}_A) \Psi^B_0(\underline{\xv}_B) \sqrt{1+ \sum_{i \in A, j \in B} \correltwo_R(\rv_{i}, \rv_j)}
\label{eq:modelwf}
\end{equation}
We regard the electrons on fragment $A$ and $B$ as localized on the fragment and ignore the contributions from antisymmetrization between the two monomers, which contributes to the energy with terms of order $\sim e^{-R}$ for large $R$. Note that the full wavefunction is still anti-symmetric under exchange of electrons within the fragments. When $R \rightarrow \infty$ we must recover the uncorrelated product $\Psi^A_0(\underline{\xv}_A) \Psi^B_0(\underline{\xv}_B)$, implying that $\correltwo_{R\to\infty}(\rv_{i}, \rv_j) \rightarrow 0$, so that $|J_R(\rv_{i},\rv_{j})|\ll 1$ for $R$ large enough. The distinguishing feature of our approach is that we now impose the constraint that for each $R$ the wavefunction yields the same $N_A$-body and $N_B$-body densities as the uncorrelated wavefunction,
\begin{align}
\int  |\Psi_R(\underline{\xv}_A, \underline{\xv}_B)|^2 \mathrm{d} \underline{\xv}_B = |\Psi^A_0(\underline{\xv}_A)|^2 \,\, \forall \,\, R, \underline{\xv}_A\label{eq:cons1A},\\
\int  |\Psi_R(\underline{\xv}_A, \underline{\xv}_B)|^2 \mathrm{d} \underline{\xv}_A = |\Psi^B_0(\underline{\xv}_B)|^2 \,\, \forall \,\, R, \underline{\xv}_B. \label{eq:cons1B}
\end{align}
Thanks to the square root ansatz with the pair correlation factor this can be simply expressed in terms of one-body density constraints,
\begin{align}
\int |\Psi^B_0(\underline{\xv}_B)|^2 J_R(
\rv_{1_A}, \rv_{1_B}) \mathrm{d} \underline{\xv}_B = \int 
\frac{\rho_0^B(\rv_{1_B})}{N_B} \correltwo_R(\rv_{1_A}, \rv_{1_B}) \mathrm{d} \rv_{1_B} = 0\, \,  \forall \,\,\rv_{1_A}, \label{eq:cons2A}\\
\int |\Psi^A_0(\underline{\xv}_A)|^2 J_R(
\rv_{1_A}, \rv_{1_B}) \mathrm{d} \underline{\xv}_A = \int 
\frac{\rho_0^A(\rv_{1_A})}{N_A} \correltwo_R(\rv_{1_A}, \rv_{1_B}) \mathrm{d} \rv_{1_A} = 0\, \,  \forall \,\,\rv_{1_B}.\label{eq:cons2B}
\end{align}
To enforce these constraints we expand the correlation function $\correltwo_R$ in terms of one-body functions $b_i^{A/B}$,
\begin{equation}
\correltwo_R(\rv_{1_A}, \rv_{1_B}) = \sum_{ij} \correlmat_{ij, R} b_i^A(\rv_{1_A}) b_j^B(\rv_{1_B}).\label{eq:expansion}
\end{equation}
The one-body density constraints of equations \eqref{eq:cons2A}-\eqref{eq:cons2B} can then be fulfilled by redefining the $b_i^{A/B}$ in terms of the functions $f_i^{A/B}$ as,
\begin{equation}
b_i^{A/B}(\rv_{1_{A/B}}) = f_i^{A/B}(\rv_{1_{A/B}}) - p_i^{A/B}, 
\end{equation}
where,
\begin{equation}
p_i^{A/B} = \int \frac{\rho_0^A(\rv_{1_{A/B}})}{N_{A/B}} f_i^{A/B}(\rv_{1_{A/B}})\mathrm{d} \rv_{1_{A/B}}, 
\end{equation}
which is simply the usual Gram-Schmidt orthogonalization procedure.

A point that deserves special attention is that the constraints of Eqs.~\eqref{eq:cons1A}-\eqref{eq:cons1B} take the simple forms of Eqs.~\eqref{eq:cons2A}-\eqref{eq:cons2B} as long as $1+J_R\ge 0$ in the full configuration space. As we will see, to leading order, we have that $J_R\propto R^{-3}$, making $|J_R|\ll 1$ for $R$ large enough, ensuring in general that this positivity constraint is fulfilled. However, by choosing functions $f_i^{A/B}(\rv)$ that increase too fast with $|\rv|$ one could violate this condition in some regions of space. In our first calculations reported in this work we have used powers $r^n$ times spherical harmonics (because of the simplicity of the integrals involved), implying that for large $n$ there could be regions of space in which $1+J_R< 0$. Nonetheless, for any maximum power $n_{max}$, we can always choose $R$ large enough to make these regions have exponentially small ($\sim e^{-R}$) effect on the energy, as they will appear only in the far tail of one (or both) of the monomer densities. Since we will solve the variational equation order by order in powers of $1/R$, these regions have formally no effect on the dispersion coefficients. In future work we will explore other choices for the $f_i^{A/B}(\rv)$.

We now turn to the computation of the matrix elements. We need to compute the expectation value of the hamiltonian, which we partition as,
\begin{align}
\hat{H} &= \hat{H}_A + \hat{H}_B+\hat{H}_{\rm int,}\\
\hat{H}_A &= \hat{T}^A+\hat{V}_{\rm ext}^A + \hat{W}^A + \hat{V}_{NN}^A = \sum_{i \in  A} \left(- \frac{1}{2} \nabla_{i}^2 - \sum_{j \in  A} \frac{Z_j}{|\mathbf{r}_i - \mathbf{R}_j|}\right)  + \sum_{i > j \in  A} \frac{1}{|\mathbf{r}_{i} - \mathbf{r}_{j}|} + \sum_{i > j \in  A} \frac{Z_i Z_j}{|\mathbf{r}_{i} - \mathbf{r}_{j}|}\\
\hat{H}_B &= \hat{T}^B+\hat{V}_{\rm ext}^B + \hat{W}^B + \hat{V}_{NN}^B =  \sum_{k \in  B} \left(- \frac{1}{2} \nabla_{k}^2 - \sum_{l \in  B} \frac{Z_l}{|\mathbf{r}_k - \mathbf{R}_l|}\right)  + \sum_{k > l \in  B} \frac{1}{|\mathbf{r}_{k} - \mathbf{r}_{l}|}+ \sum_{k > l \in  B} \frac{Z_k Z_l}{|\mathbf{r}_{k} - \mathbf{r}_{l}|}\\
\hat{H}_{\rm int} &= \sum_{i \in  A, k \in  B} \frac{1}{|\mathbf{r}_i - \mathbf{r}_k|} -  \sum_{i \in  A, k \in  B} \frac{Z_k}{|\mathbf{r}_i - \mathbf{R}_k|} - \sum_{i \in  A, k \in  B} \frac{Z_i}{|\mathbf{r}_k - \mathbf{R}_i|} + \sum_{i \in  A, k \in  B} \frac{Z_i Z_k}{|\mathbf{R}_i - \mathbf{R}_k|} =: \sum_{i \in A, k \in B} w_{\rm int}(\mathbf{r}_i, \mathbf{r}_k),
\end{align}
where we have indicated with $\mathbf{R}_i$ nuclear coordinates, with $\mathbf{r}_i$ electron coordinates, and with $Z_i$ the nuclear charges.
Due to the $N_A$/$N_B$-body density conservation the expectation values of the external potential $\hat{V}_\mathrm{ext}^{A/B}$ and electron-electron interaction inside each fragment $\hat{W}^{A/B}$ are not changed by our wavefunction with respect to their isolated ground state values. More explicitly, for the expectation values of the external potential and electron-electron interaction on monomer $A$,
\begin{align}
\langle \Psi_R | \hat{V}_\mathrm{ext}^A | \Psi_R \rangle &= N_A \int  |\Psi_R(\underline{\xv}_A, \underline{\xv}_B)|^2 v_{\rm ext}^A(\rv_{1_A}) \mathrm{d} \underline{\xv}_A \mathrm{d} \underline{\xv}_B = N_A \int |\Psi^A_0(\underline{\xv}_A)|^2 v_{\rm ext}^A(\rv_{1_A}) \mathrm{d} \underline{\xv}_A = \langle \Psi^A_0 | \hat{V}_\mathrm{ext}^A | \Psi^A_0 \rangle, \\
\langle \Psi_R | \hat{W}^A | \Psi_R \rangle &= \frac{1}{2}\int  |\Psi_R(\underline{\xv}_A, \underline{\xv}_B)|^2 \frac{N_A(N_A-1) }{|\rv_{1_A}- \rv_{2_A}|} \mathrm{d} \underline{\xv}_A \mathrm{d} \underline{\xv}_B = \frac{1}{2} \int |\Psi^A_0(\underline{\xv}_A)|^2 \frac{N_A(N_A-1) }{|\rv_{1_A}- \rv_{2_A}|} \mathrm{d} \underline{\xv}_A = \langle \Psi^A_0 | \hat{W}^A | \Psi^A_0 \rangle ,
\end{align}
and the same for monomer $B$.

The expectation values of the kinetic energy operator $\hat{T}^{A/B}$ and the expectation value of the interaction energy operator $\hat{H}_\mathrm{int}$ can be split into a non-interacting part (where interaction here refers to the monomer-monomer one), independent of the variational parameters $c_{ij}$, and a correlation part dependent on $c_{ij}$. The expectation value of the interaction energy operator $\hat{H}_\mathrm{int}$ is simply,
\begin{align}
\langle \Psi_R | \hat{H}_\mathrm{int} | \Psi_R \rangle &= N_A N_B \int w_\mathrm{int}(\rv_{1_A},\rv_{1_B}) |\Psi^A_0(\underline{\xv}_A)|^2 |\Psi^B_0(\underline{\xv}_B)|^2 \left(1+ \sum_{i \in A, j \in B} \correltwo_R(\rv_{i}, \rv_j)\right) \mathrm{d} \underline{\xv}_A  \mathrm{d} \underline{\xv}_B  \\
&= \langle \Psi_0^A \Psi_0^B| \hat{H}_\mathrm{int} | \Psi_0^A\Psi_0^B \rangle + N_A N_B \sum_{i \in A, j \in B} \int w_\mathrm{int}(\rv_{1_A},\rv_{1_B}) |\Psi^A_0(\underline{\xv}_A)|^2 |\Psi^B_0(\underline{\xv}_B)|^2  \correltwo_R(\rv_{i}, \rv_j) \mathrm{d} \underline{\xv}_A  \mathrm{d} \underline{\xv}_B  \\
&=: W_{\mathrm{int},0}(R) + W_\mathrm{ \mathrm{int}, c}(\{c_{ij}\},R),
\end{align}
where $W_{\mathrm{int},0}(R)$ is the familiar electrostatic energy due to the interaction between the unperturbed densities of the two monomers.
The summation in $W_\mathrm{ \mathrm{int}, c}(\{c_{ij}\},R)$ can be split into four contributions: $i = \rv_{1_A}, j= \rv_{1_B}$ with a multiplicity of one,  $i = \rv_{1_A}, j= \rv_{2_B}$ with a multiplicity of $(N_B-1)$, $i = \rv_{2_A}, j= \rv_{1_B}$ with a multiplicity of $(N_A-1)$ and $i=\rv_{2_A}, j= \rv_{2_B}$ with a multiplicity of $(N_A-1)(N_B-1)$, yielding 
\begin{align}
W_\mathrm{ \mathrm{int}, c}(\{c_{ij}\},R) =& \int w_\mathrm{int}(\rv_{1_A},\rv_{1_B}) \rho_0^A(\rv_{1_A})\rho_0^B(\rv_{1_B}) \correltwo_R(\rv_{1_A}, \rv_{1_B}) \mathrm{d} \rv_{1_A}  \mathrm{d} \rv_{1_B}   \\ &+ \int w_\mathrm{int}(\rv_{1_A},\rv_{1_B}) P_0^A(\rv_{1_A}, \rv_{2_A})\rho_0^B(\rv_{1_B}) \correltwo_R(\rv_{2_A}, \rv_{1_B}) \mathrm{d} \rv_{1_A}  \mathrm{d} \rv_{2_A} \mathrm{d} \rv_{1_B} \nonumber    \\
&+ \int w_\mathrm{int}(\rv_{1_A},\rv_{1_B}) \rho_0^A(\rv_{1_A}) P_0^B(\rv_{1_B}, \rv_{2_B}) \correltwo_R(\rv_{1_A}, \rv_{2_B}) \mathrm{d} \rv_{1_A} \mathrm{d} \rv_{1_B} \mathrm{d} \rv_{2_B}  \nonumber \\ &+ \int w_\mathrm{int}(\rv_{1_A},\rv_{1_B}) P_0^A(\rv_{1_A}, \rv_{2_A}) P_0^B(\rv_{1_B}, \rv_{2_B}) \correltwo_R(\rv_{2_A}, \rv_{2_B}) \mathrm{d} \rv_{1_A} \mathrm{d} \rv_{2_A} \mathrm{d} \rv_{1_B} \mathrm{d} \rv_{2_B},  \nonumber
\end{align}
where $\rho_0^{A(B)}(\rv)$ are the ground-state monomer densities and $P_0^{A(B)}(\rv_1,\rv_2)$ are their pair densities (the diagonal of the two-body reduced density matrix).
Introducing the expansion of $\correltwo_R$ into one-body states we can rewrite this expectation value as
\begin{align}
W_\mathrm{ \mathrm{int}, c}(\{c_{ij}\},R) =& \sum_{ij} c_{ij} w_{ij}(R),\\
w_{ij} =& \int w_\mathrm{int}(\rv_{1_A},\rv_{1_B}) \rho_0^A(\rv_{1_A})\rho_0^B(\rv_{1_B}) b^A_i(\mathbf{r}_{1_A}) b^B_j(\mathbf{r}_{1_B}) \mathrm{d} \rv_{1_A}  \mathrm{d} \rv_{1_B}  \\ &+ \int w_\mathrm{int}(\rv_{1_A},\rv_{1_B}) P_0^A(\rv_{1_A}, \rv_{2_A})\rho_0^B(\rv_{1_B}) b^A_i(\mathbf{r}_{2_A}) b^B_j(\mathbf{r}_{1_B}) \mathrm{d} \rv_{1_A}  \mathrm{d} \rv_{2_A} \mathrm{d} \rv_{1_B}  \nonumber  \\
&+ \int w_\mathrm{int}(\rv_{1_A},\rv_{1_B}) \rho_0^A(\rv_{1_A}) P_0^B(\rv_{1_B}, \rv_{2_B}) b^A_i(\mathbf{r}_{1_A}) b^B_j(\mathbf{r}_{2_B}) \mathrm{d} \rv_{1_A} \mathrm{d} \rv_{1_B} \mathrm{d} \rv_{2_B}  \nonumber \\ &+ \int w_\mathrm{int}(\rv_{1_A},\rv_{1_B}) P_0^A(\rv_{1_A}, \rv_{2_A}) P_0^B(\rv_{1_B}, \rv_{2_B}) b^A_i(\mathbf{r}_{2_A}) b^B_j(\mathbf{r}_{2_B}) \mathrm{d} \rv_{1_A} \mathrm{d} \rv_{2_A} \mathrm{d} \rv_{1_B} \mathrm{d} \rv_{2_B}   \nonumber
\end{align}
If we now perform the usual multipole expansion of $w_{\rm int}(\rv_A,\rv_B)$, the interaction becomes separable in the individual moments, and the expression can be further simplified at each order in $1/R$. For example, for the dipole-dipole coupling we can write,
\begin{align}
W_\mathrm{ \mathrm{int}, c}(\{c_{ij}\},R)  = \frac{1}{R^3} \sum_{ij} c_{ij} \sum_{e,f=x,y,z} h_{ef} (d_{i,e}^A + D_{i,e}^A) (d_{j,f}^B+D_{j,f}^B),
\end{align}
where $h_{ef}$ depends on the relative orientation of the two monomers and we have defined,
\begin{align}
 d_{i,e}^A &= \int \mathrm{d} \mathbf{r}_{1_A} \rho_0^A(\mathbf{r}_{1_A})b_i^A(\mathbf{r}_{1_A}) e_{1_A}, \\
D_{i,e}^A &= \int \mathrm{d} \mathbf{r}_{1_A} \int \mathrm{d} \mathbf{r}_{2_A} P_0^A(\mathbf{r}_{1_A}, \mathbf{r}_{2_A}) b_i^A(\mathbf{r}_{2_A}) e_{1_A}.
\end{align}

Obtaining the kinetic energy is slightly more involved. We need to compute (and similarly for $B$):
\begin{equation}
\langle \Psi_R | \hat{T}^A | \Psi_R \rangle = \frac{N_A}{2} \int | \nabla_{\rv_{1_A}} \Psi_R(\underline{\xv}_A, \underline{\xv}_B)|^2 \mathrm{d} \underline{\xv}_A \mathrm{d} \underline{\xv}_B.
\end{equation}
We start with the expression for the gradient, 
\begin{equation}
\nabla_{\rv_{1_A}} \Psi_R(\underline{\xv}_A, \underline{\xv}_B) = \Psi^B_0(\underline{\xv}_B) \left(\sqrt{1+ \sum_{i \in A, j \in B} \correltwo_R(\rv_{i}, \rv_j)} \nabla_{\rv_{1_A}} \Psi^A_0(\underline{\xv}_A)+ \Psi^A_0(\underline{\xv}_A) \frac{\sum_{k \in B} \nabla_{\rv_{1_A}} \correltwo_R(\rv_{1_A}, \rv_k)}{2\sqrt{1+ \sum_{i \in A, j \in B} \correltwo_R(\rv_{i}, \rv_j)}} \right), 
\end{equation}
leading to the following expression for its square,
\begin{equation}
\begin{split}
|\nabla_{\rv_{1_A}} \Psi_R(\underline{\xv}_A, \underline{\xv}_B)|^2 = |\Psi^B_0(\underline{\xv}_B)|^2 \Big( \big(1+ \sum_{i \in A, j \in B} \correltwo_R(\rv_{i}, \rv_j)\big) |\nabla_{\rv_{1_A}} \Psi^A_0(\underline{\xv}_A)|^2 \\ + |\Psi^A_0(\underline{\xv}_A)|^2 \frac{\sum_{k \in B, l \in B} \nabla_{\rv_{1_A}} \correltwo_R(\rv_{1_A}, \rv_k) \cdot \nabla_{\rv_{1_A}} \correltwo_R(\rv_{1_A}, \rv_l)}{4(1+ \sum_{i \in A, j \in B} \correltwo_R(\rv_{i}, \rv_j))} +  \Psi^A_0(\underline{\xv}_A) \sum_{k \in B} \nabla_{\rv_{1_A}} \correltwo_R(\rv_{1_A}, \rv_k) \cdot\nabla_{\rv_{1_A}} \Psi^A_0(\underline{\xv}_A)\Big).
\end{split}
\end{equation}
The first term only contributes the kinetic energy of the non-interacting fragments due to the density constraint:
\begin{equation}
\frac{N_A}{2} \int |\nabla_{\rv_{1_A}} \Psi^A_0(\underline{\xv}_A)|^2 \mathrm{d} \underline{\xv}_{A} \int|\Psi^B_0(\underline{\xv}_B)|^2 \Big(1+ \sum_{i \in A, j \in B} \correltwo_R(\rv_{i}, \rv_j)\Big) \mathrm{d} \underline{\xv}_{B} = \frac{N_A}{2} \int |\nabla_{\rv_{1_A}} \Psi^A_0(\underline{\xv}_A)|^2 \mathrm{d} \underline{\xv}_{A} = T^A_0.
\end{equation}
The third term is zero as a result of the density constraint,
\begin{equation}
 \frac{N_A}{2}\int \mathrm{d} \underline{\xv}_A \Psi^A_0(\underline{\xv}_A) \nabla_{\rv_{1_A}} \Psi^A_0(\underline{\xv}_A) \cdot  \nabla_{\rv_{1_A}} \int \mathrm{d} \underline{\xv}_B \sum_{k \in B} \correltwo_R(\rv_{1_A}, \rv_k)  |\Psi^B_0(\underline{\xv}_B)|^2 = 0.
\end{equation}
The second term is the desired kinetic correlation energy,
\begin{equation}
T_c^A = \frac{N_A}{8} \sum_{k \in B, l \in B} \int \mathrm{d} \underline{\xv}_A \mathrm{d} \underline{\xv}_B |\Psi^A_0(\underline{\xv}_A)|^2 |\Psi^B_0(\underline{\xv}_B)|^2 \frac{ \nabla_{\rv_{1_A}} \correltwo_R(\rv_{1_A}, \rv_k) \cdot \nabla_{\rv_{1_A}} \correltwo_R(\rv_{1_A}, \rv_l)}{1+ \sum_{i \in A, j \in B} \correltwo_R(\rv_{i}, \rv_j)}. \label{eq:fullkincorrelation}
\end{equation}
Notice that the change in kinetic energy becomes, to leading order, {\em quadratic} in $J_R$ only because of the density constraint, which makes the terms linear in $J_R$ vanish. \\ To compute dispersion coefficients we now use the fact that we know that $c_{ij}\ll 1$ for large $R$, so we perform a series expansion in $c_{ij}$ of the correlation kinetic energy and truncate it after the first non-zero term, which amounts to simply getting rid of the denominator in equation \eqref{eq:fullkincorrelation}. This corresponds to performing second-order perturbation theory with $\hat{H}_{\rm int}$ as the perturbation. Second-order perturbation theory provides the entire contribution to the dispersion coefficients $C_6$, $C_8$ and $C_{10}$. Within this approximation (which is exact up to $C_{10}$) we write
\begin{equation}
T_c^A \approx \frac{N_A}{8} \sum_{k \in B, l \in B} \int \mathrm{d} \underline{\xv}_A \mathrm{d} \underline{\xv}_B |\Psi^A_0(\underline{\xv}_A)|^2 | \Psi^B_0(\underline{\xv}_B)|^2\nabla_{\rv_{1_A}} \correltwo_R(\rv_{1_A}, \rv_k) \cdot \nabla_{\rv_{1_A}} \correltwo_R(\rv_{1_A}, \rv_l).
\end{equation}
We can again isolate the different terms in the sum. Either $l=k$ (multiplicity $N_B$) or $l \neq k$ (multiplicity $N_B (N_B-1)$).  
\begin{equation}
\begin{split}
T_c^A =& \frac{1}{8} \int \mathrm{d} \rv_{1_A} \mathrm{d} \rv_{1_B} \rho_0^A(\rv_{1_A}) \rho_0^B(\rv_{1_B}) |\nabla_{\rv_{1_A}} \correltwo_R(\rv_{1_A}, \rv_{1_B})|^2\\& + \frac{1}{8} \int \mathrm{d} \rv_{1_A} \mathrm{d} \rv_{1_B} \mathrm{d} \rv_{2_B} \rho_0^A(\rv_{1_A}) P_0^B(\rv_{1_B}, \rv_{2_B}) \nabla_{\rv_{1_A}} \correltwo_R(\rv_{1_A}, \rv_{1_B}) \cdot \nabla_{\rv_{1_A}} \correltwo_R(\rv_{1_A}, \rv_{2_B})
\end{split}
\end{equation}
Obtaining the analogous expression for fragment $B$ and expanding $J$ in the one-particle basis we can write the kinetic energy as,
\begin{equation}
T_c^A(\{c_{ij}\}) + T_c^B(\{c_{ij}\}) =  \frac{1}{2} c_{ij} c_{kl} L_{ij, kl}, 
\end{equation}
where we have defined $L_{ij,kl}$ as,
\begin{subequations}
\label{eq:Ldefinition}
\begin{align}
L_{ij, kl}&= \frac{1}{4} ( \tau_{ik}^A (S_{jl}^B+P_{jl}^B) + \tau_{jl}^B (S_{ik}^A+P_{ik}^A)),\\
\tau_{ij} &= \int \mathrm{d} \rv \rho(\rv) \nabla_\rv b_i(\rv) \cdot \nabla_\rv b_j(\rv),\\
S_{ij}&= \int \mathrm{d} \rv \rho(\rv) b_i(\rv) b_j(\rv),\\
P_{ij}&= \int \mathrm{d} \rv_1 \mathrm{d} \rv_2 P(\rv_1, \rv_2) b_i(\rv_1) b_j(\rv_2).
\end{align}
\end{subequations}
Adding all the contributions together we obtain a minimization problem reminiscent of the Levy constrained search \cite{Lev-PNAS-79} for the universal functional of density functional theory (DFT), consisting of a minimization over wavefunctions yielding the ground-state many-particle densities of the monomers.  The energy of the combined systems $E(R)$ is given by:
\begin{align}
E(R) &= E_0^A+E_0^B + E_{\rm int}(R),\\
E_{\rm int}(R)&= \min_{\Psi \rightarrow |\Psi_0^A|^2, |\Psi_0^B|^2} \langle \Psi | \hat{H}_{\rm int}+\hat{T}^A+\hat{T}^B | \Psi \rangle  - T_0^A - T_0^B =\nonumber \\
&=  W_{\mathrm{int},0}(R)+\min_{\{c_{ij}\}} (W_\mathrm{ \rm \mathrm{int}, c}(\{c_{ij}\},R) + T^A_c( \{c_{ij}\} ) + T^B_c(\{c_{ij}\}),
\end{align}
where $E_0^{A/B}$ are the ground-state energies of the two monomers. $E_{\rm int}(R)$ is the interaction energy and is composed of a potential energy contribution and a kinetic contribution. $W_{\mathrm{int},0}(R)$ is the classical electrostatic contribution, which is the expectation value of the interaction hamiltonian on the ground-state of the monomers. $W_\mathrm{ \mathrm{int}, c}(\{c_{ij}\},R)$ is the potential energy correlation contribution to the interaction-energy.  $T^{A/B}_c( \{c_{ij}\} )$ is the kinetic correlation contribution.\\ 
Our functional $E_{\rm int}$ to second-order in $c_{ij}$ can be written as:
\begin{equation}
E_{\rm int}(R) = \min_{\{c_{ij}\}} \left(\sum_{ij} c_{ij} w_{ij}(R)  + \frac{1}{2} \sum_{ijkl} c_{ij} c_{kl} L_{ij, kl}\right)
\end{equation}
This is a quadratic expression $c_{ij}$, such that minimization yields a linear system,
\begin{equation}
L \mathbf{c} = -  \mathbf{w}(R).
\end{equation}
 Note that the vector $\mathbf{w}$ has one contracted index $I=ij$, while the matrix $L$ has two contracted indices $I=ij$ and $K=kl$. Plugging in the solution we obtain,
\begin{equation}
E_{ \rm int}(R)= - \frac{1}{2} \mathbf{w}(R)^\intercal L^{-1} \mathbf{w}(R)
\end{equation}
For example for the dipole-dipole interaction, $\mathbf{w}(R) \propto R^{-3}$ and therefore we obtain that $E_{\rm int} \propto R^{-6}$. In practice one does not want to  solve the (high-dimensional) linear problem, which can be highly simplified. Looking back at equations \eqref{eq:Ldefinition} we can introduce a unitary transformation among the functions $b_i$, such that $S_{ik}^A+P_{ik}^A = \delta_{ik}$ and similar for monomer $B$. This can be achieved with a simple L\"owdin orthogonalization. In this case the linear equation can be written as:
\begin{align}
\sum_{kl} L_{ij,kl} c_{kl} &= -w_{ij} \\
\sum_{kl}\frac{1}{4} (\tau_{ik}^A \delta_{jl}^B + \delta_{ik}^A \tau_{jl}^B) c_{kl} &= -w_{ij}\\
\sum_{k}\tau_{ik}^A c_{kj} + \sum_{l} \tau_{jl}^B c_{il} &= -4 w_{ij}, 
\end{align}
becoming a Sylvester equation \cite{BarSte-ACM-72}, in matrix form:
\begin{equation}
\tau^A c + c (\tau^B)^\intercal = - 4 w,
\end{equation}
which can be solved in an efficient way, as described in \cite{BarSte-ACM-72}.

\section{Two atoms}
In the case of two atoms we may introduce further simplifications due to isotropy and it is very convenient to work with the expansion in spherical harmonics.  For the higher order $C_{2n}$ we develop the expansion of the interaction potential to arbitrary order, placing the two atoms aligned on the z-axis, in which case the interaction potential is,
 \begin{equation}
 \begin{split}
w_\mathrm{int}(\rv_{1_A}, \rv_{1_B})= \sum_{l_A=1}^\infty \sum_{l_B=1}^\infty (-1)^{l_B} \binom{2l_A + 2 l_B}{2 l_A}^{1/2} R^{-l_A-l_B-1} \sum_{m=\max(-l_A,-l_B)}^{\min(l_a,l_B)} \hat{q}_{m}^{A,l_A} \hat{q}_{-m}^{B,l_B} \langle l_A,m;l_B,-m|l_A+l_B, 0\rangle,
 \end{split}
 \end{equation}
 where
 \begin{equation}
\hat{q}_{m}^{A, l}(\mathbf{r}) = r_{1}^{l} \sqrt{\frac{4 \pi}{2l+1}} Y_{l, m}(\theta, \phi),
 \end{equation}
 yielding the general expression
 \begin{align}
W_\mathrm{ \rm \mathrm{int}, c}(\{c_{ij}\},R)=&\sum_{ij} c_{ij} \sum_{l_A=1}^\infty \sum_{l_B=1}^\infty (-1)^{l_B} \binom{2l_A + 2 l_B}{2 l_A}^{1/2} R^{-l_A-l_B-1} \\ & \sum_{m=\max(-l_A,-l_B)}^{\min(l_a,l_B)} (q_{i, m}^{A,l_A}+Q_{i, m}^{A,l_A})(q_{j,-m}^{B,l_B}+Q_{j, -m}^{B,l_B}) \langle l_A,m;l_B,-m|l_A+l_B, 0\rangle \nonumber,
\end{align}
with
\begin{align}
q_{i, m}^{l} &= \int \rho(\mathbf{r}) \hat{q}_{m}^{l} b_i(\mathbf{r})  \mathrm{d} \mathbf{r},\\
Q_{i, m}^{l} &= \int P(\mathbf{r}_1, \mathbf{r}_2) \hat{q}_{m}^{l}(\mathbf{r}_1) b_i(\mathbf{r}_2)  \mathrm{d} \mathbf{r}_1 \mathrm{d} \mathbf{r}_2.
\end{align}
We restrict our $f_i$  and thus $b_i$ (since in the atomic case $p_i = 0$) to:
\begin{equation}
b_i(\mathbf{r}) = r^{n_i} Y_{l_i, m_i}(\theta, \phi), 
\end{equation}
in which case only $q_{i, -m_i}^{A/B,l_i}$ and $Q_{i, -m_i}^{A/B,l_i}$  are non-zero and, furthermore, the value is independent of $m_i$.  We therefore can perform the calculation as such for each pair of values $l_i$ and $l_j$. 
Defining,
\begin{equation}
w_{ij}^{l_a,l_b} = q_{i, 0}^{A/B, l_a} q_{j,0}^{A/B, l_b},
\end{equation}
we can express the contribution from the induced $l_a$-pole interacting with the induced $l_b$-pole:
\begin{equation}
\mathcal{Q}_{l_a, l_b} = \frac{((l_a+l_b)!)^2 (2 \, _3F_2(1,-l_a,-l_b;l_a+1,l_b+1;1)-1 )}{2(l_a!)^2 (l_b!)^2} {\mathbf{w}^{l_a,l_b}}^\intercal L^{-1} \mathbf{w}^{l_a,l_b}, 
\end{equation}
Summing the different contributions to each of the coefficients, we obtain:
\begin{equation}
C_{2n} = \sum_{l=1}^{n-1}  \mathcal{Q}_{l, n-l},
\end{equation}
note that we only obtain the second-order dispersion effects due to the truncation of the kinetic energy after the second order in $c_{ij}$. This means that we obtain the exact $C_{6}$, $C_8$ and $C_{10}$, but we only recover part of (for example) $C_{12}$, which also has a contribution from fourth-order perturbation theory. We compare our results to the second-order dispersion coefficients from Thakkar\cite{Tha-JCP-88} in Table \ref{table:C30}. A rather curious observation is that the convergence seems to be better for higher multipoles.

\begin{table}
\begin{tabular}{|c|c|c|c|}
\hline
Coefficient &  Thakkar \cite{Tha-JCP-88} & This work (20 states) & This work (30 states)\\
\hline
6 & $6.4990267054058405\times 10^0$ & $6.4990267054058\mathbf{366}\times 10^0$ & $6.4990267054058\mathbf{393}\times 10^0$\\
8 & $1.2439908358362235\times 10^2$ & $1.243990835836223\mathbf{3}\times 10^2$ & $1.243990835836223\mathbf{4}\times 10^2$\\
10 & $3.2858284149674217\times 10^3$ & $3.285828414967421\mathbf{6}\times 10^3$ & $3.2858284149674217\times 10^3$\\
12 & $1.2148602089686110\times 10^5$ & $1.2148602089686110\times 10^5$& ibid. \\
14 & $6.0607726891921249\times 10^6$ & $6.0607726891921249\times 10^6$ & ibid.\\
16 & $3.9375063939991797\times 10^8$ & $3.9375063939991797\times 10^8$& ibid. \\
18 &$3.2342187158493778\times 10^{10}$ & $3.2342187158493778\times 10^{10}$ & ibid.\\
20 & $3.2785734404166217\times 10^{12}$ & $3.2785734404166217\times 10^{12}$& ibid. \\
22 & $4.0210828476853598\times 10^{14}$ & $4.0210828476853598\times 10^{14}$& ibid. \\
24 & $5.8689963345599573\times 10^{16}$ & $5.8689963345599573\times 10^{16}$& ibid. \\
26 & $1.0052949933362942\times 10^{19}$ & $1.0052949933362942\times 10^{19}$ & ibid.\\
28 & $1.9969449408875758\times 10^{21}$ & $1.9969449408875758\times 10^{21}$ & ibid.\\
30 & $4.5532888666347351 \times 10^{23}$ & $4.5532888666347351\times 10^{23}$& ibid. \\
\hline
\end{tabular}
\caption{Second-order dispersion coefficients obtained using our method (with 20 and 30 $b$ states). Digits in bold differ from the result of Thakkar\cite{Tha-JCP-88}.\label{table:C30}}
\end{table}

\bibliography{../../RsearchGroupAdminstration/bib_clean.bib,../references/dispersion_bib.bib}